# A General and Modular Approach to Solid-state Integration and Readout of Zero-dimensional Quantum Systems

*Marzieh Kavand,[1,6] Zoe Phillips,[1,‡] William H. Koll,[1,‡], Morgan Hamilton,[1,‡] Ethel Perez-Hoyos,[1,†] Rianna Greer,[2] Ferdous Ara,[1,†] Daniel Pharis,[1,†] Kian Maleki,[3] Mingyu Xu,[4] Takashi Taniguchi,[5] Paul Canfield,[4] Michael E. Flatté,[3,7] Danna E. Freedman,[2] Jay Gupta,[1] Ezekiel Johnston-Halperin[1*]*

[1]Department of Physics, The Ohio State University, Columbus, Ohio, 43210, USA.

[2]Department of Chemistry, Massachusetts Institute of Technology, Cambridge, Massachusetts 02139, USA. [3]Department of Physics and Astronomy, University of Iowa, Iowa City, Iowa, 52242, USA.

[4]Ames National Laboratory and Department of Physics and Astronomy, Iowa State University, Ames, Iowa, 50010, USA.

[5]International Center for Materials Nanoarchitectonics (MANA), National Institute for Materials Science, 1-1 Namiki Tsukuba Ibaraki 305-0044, Japan.

[6]Department of Physics and Astronomy, University of Alabama, Tuscaloosa, Alabama, 35487, USA.

[7]Department of Applied Physics and Science Education, Eindhoven University of Technology, 6500 MB Eindhoven, The Netherlands.

**ABSTRACT:** Here, we present an all-electrical readout mechanism for quasi-0D quantum states (0D-QS) such as point defects, adatoms and molecules, that is modular and general, providing an approach that is amenable to scaling and integration with other solid-state quantum technologies. Our approach relies on the creation of high-quality tunnel junctions via the mechanical exfoliation and stacking of multi-layer graphene (MLG) and hexagonal boron nitride (hBN) to encapsulate the target system in an MLG/hBN/0D-QS/hBN/MLG heterostructure. This structure allows for all-electronic spectroscopy and readout of candidate systems through a combination of coulomb and spin-blockade. As a proof of principle, we demonstrate electronic tunneling spectroscopy of point defects in hBN and the molecular qubit vanadyl phthalocyanine (VOPc). Our approach demonstrates a new pathway for the incorporation of molecules and atomic defects into solid-state quantum devices and circuits along with a readout scheme that does not rely on highly-constrained optical processes for photonic readout.

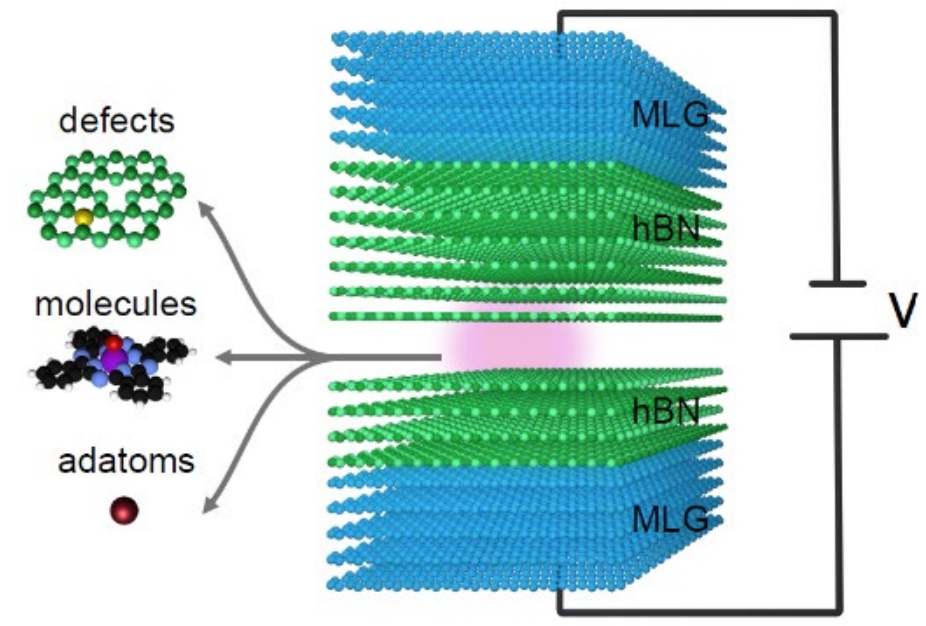

Electronic spectroscopy of zero-dimensional (0D) quantum systems, including point defects in solids[1–3], atomic states[4,5], and small molecules[6–8], is a critical tool for developing a fundamental understanding of these systems, with applications ranging from solid-state[9] and molecular[10,11] materials development to emerging technologies rooted in quantum information science[12] (QIS). In particular, the atomic control and flexibility of design afforded by molecular systems provides a particularly attractive approach to the precision engineering of quantum states required for QIS applications. However, these systems are typically not compatible with the process conditions associated with related solid-state quantum technologies. For example, device-based approaches that rely on embedding these systems within a solid-state tunnel junction (TJ) are not broadly applicable and generally degrade the molecular system, while scanning tunneling spectroscopy (STS) is relatively benign and has demonstrated atomic-scale sensitivity but is not easily scalable for extensive studies or applications. At the same time, lateral molecular devices[13–16] face challenges associated with low yield due to difficulties in establishing reliable electrical contact and deterministic fabrication. This leaves a substantial gap in our ability to study these 0D systems and integrate them into existing quantum technologies.

In addressing this challenge, there are several key constraints to consider: (i) the relevant quantum properties should be preserved in the integrated state, (ii) the electronic response of the integrated device should allow for readout of the relevant quantum properties, (iii) the quantum system should be protected from fabrication processes, (iv) the device architecture should be compatible with existing quantum technologies. Mechanical exfoliation and stacking of 2D materials provides a low-temperature and non-reactive approach to building modular tunnel junctions with the potential to satisfy these criteria. The mechanical nature of 2D exfoliation and stacking provides facile encapsulation of quantum systems between 2D layers, including molecules, adatoms, and quantum point defects (QPD), that avoids the high temperatures and reactive chemistries typically involved in traditional fabrication processes. Further, exfoliation and stacking provides a facile route to the fabrication of high-quality tunnel junctions with atomically precise control of barrier thickness and strongly suppressed pinhole formation.

Here, we demonstrate the power and modularity of this 2D exfoliation and encapsulation strategy by generating high quality tunnel junctions comprised of both QPDs and molecular quantum systems in the form of point defects in hexagonal boron nitride (hBN) and the molecular qubit vanadyl phthalocyanine (VOPc), a spin 1/2 molecular qubit that has demonstrated long coherence times in an isostructural diamagnetic host (up to 1 μs at room temperature[8]), respectively. The differential conductance (dI/dV) of these devices reveals the underlying electronic structure, and in the case of VOPc devices, we confirm both qualitative and quantitative agreement with scanning-tunneling spectroscopy (STS) measurements and the theory of VOPc electronic structure.[17] These devices are fully compatible with existing solid-state quantum technologies (e.g., silicon quantum dots[18] and superconducting qubits[19]) and experimental protocols[20–22], and our fabrication protocols are compatible with monolithic integration of microwave components. These results therefore demonstrate a modular and general approach to the investigation and manipulation of novel quantum systems that will allow for a more systematic approach to the design and manufacturing of new quantum technologies.

For example, the motion of charge through a 0D structure involves, at least temporarily, an initial and/or final state that possesses spin; study of this transport, especially in small magnetic fields, has been used to reveal quantum-technology-relevant quantities such as low-temperature spin coherence times in individual quantum dots[23,24], and at room temperature the hyperfine fields and exchange interactions of small numbers of defects[25–27]. In addition, prior work in related tunnel junctions in the Coulomb blockade[28–30] regime has demonstrated the potential for high-fidelity quantum readout[31–34]. Finally, this approach leverages a large body of prior work in integrating 2D materials into existing quantum architectures such as on-chip microwave resonators[35], transmons[36], and spin qubits[18,33].

Figure 1a shows a schematic illustration of one such hBN-based 2D tunnel junction (2D-TJ) device (see the Supporting Information for a more detailed discussion) with graphitic contacts and gold electrodes, while Figure 1b shows an optical image of a representative hBN-TJ device with an hBN tunnel barrier of 3.01±0.66 nm and an active area of 3.1 μm$^2$). The active area of the device is stacked with mechanically exfoliated hBN and graphite using well-established dry transfer techniques[37], wherein the graphitic contacts consist of multilayer graphene (MLG) of typical thickness 6.5±3.5 nm, well into the bulk graphitic regime. To achieve the desired tunneling conditions, the insulating hBN layer is selected to be between 1.5 nm and 3.5 nm, confirmed by atomic force microscopy (AFM) (see Figure S1a). The thickness of the hBN barrier dictates the current of the nonlinear tunneling response, ranging from tens of nanoamps to a few microamps in this regime. Gold electrodes are deposited onto completed stacks, with care taken to avoid extending the metal electrodes over the active area of the device in order to suppress metal migration and the creation of parasitic current pathways.

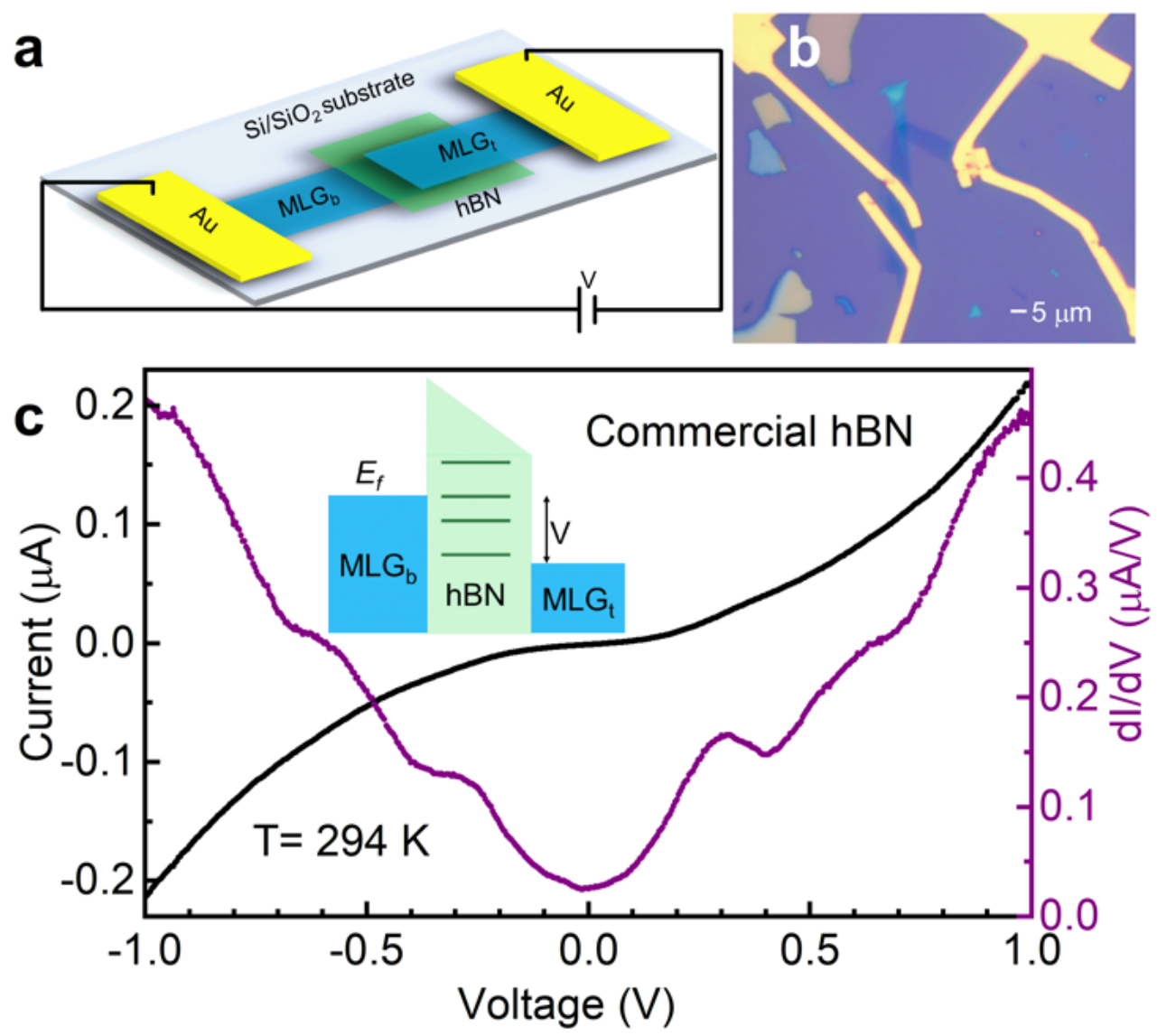

**Figure 1.** Electrical characterization of a hBN tunnel junction (hBN-TJ) device. (a) Schematic of the hBN-TJ device structure. The 2D heterostructure features multi-layer graphene (MLG) contacts to a hBN barrier layer. (b) Optical image of a hBN-TJ device. (c) I-V, black curve, and dI/dV, purple curve, at room temperature. Plateaus in I-V and broad peaks in dI/dV are due to the resonant tunneling caused by native defect states in the hBN barrier. Inset, a schematic of the energy diagram of the TJ device. Discrete lines represent the energy states of the native defects within the hBN bandgap.

Initial device characterization is performed at room temperature, as shown in Fig.1c for a device constructed with hBN sourced commercially from HQ Graphene (thickness of 4±0.5 nm). A schematic energy diagram is shown in the inset. The black curve shows the tunneling current through the device while sweeping the bias voltage (I-V response), revealing a contribution from direct tunneling through the barrier and additional tunneling processes within the device (e.g., the plateau at a source-drain bias of ±0.30 V). To highlight these additional processes, synchronous measurement of dI/dV is performed using frequency modulation and homodyne detection with a lock-in amplifier, as represented by the purple curve in Figure 1c. These data show both direct tunneling and broad peaks that appear at, for ex-ample, ±0.30 V and ± 0.60 V, suggesting the presence of resonant tunneling processes through quantum states with energies lying within the bandgap of the hBN.

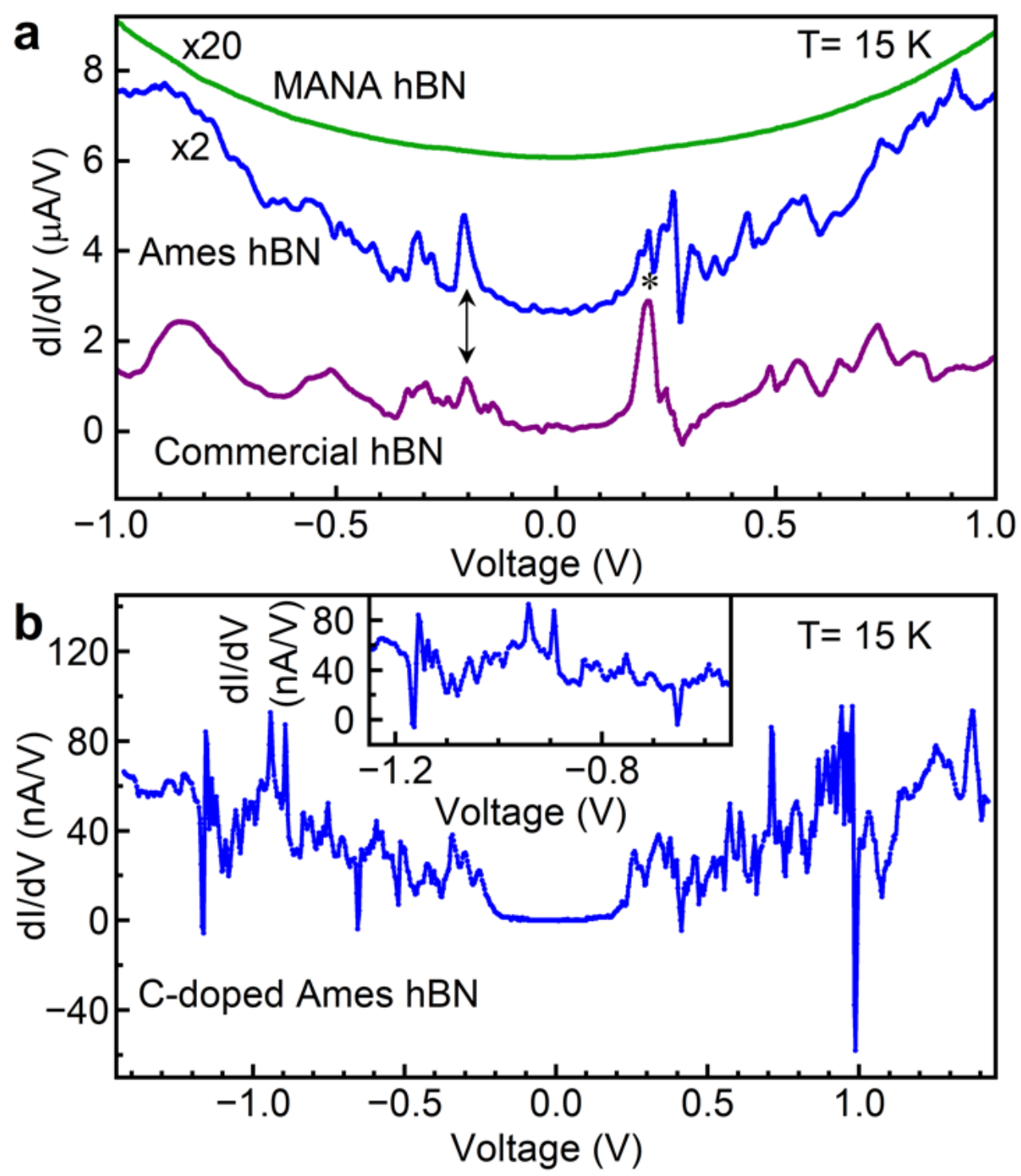

**Figure 2.** Tunneling spectroscopy of quantum point defects in hBN. (a) dI/dV of the hBN tunnel junction (hBN-TJ) devices at 15 K for three sources of pure hBN: commercial hBN from HQ Graphene, Ames hBN, and MANA hBN. Tunneling spectrum of the MANA hBN device reveals no significant resonant tunneling peaks (See Figure S2). (b) dI/dV of carbon-related defects in Ames hBN at 15 K. Inset, a zoomed in dI/dV around -1 V, indicating the sharpness of the atomic-like resonant peaks.

To resolve these resonances, additional measurements were performed at a temperature of 15 K (purple curve in Figure 2a) for a similar device (hBN thickness of 3.01±0.66 nm), and the reduction in thermal broadening reveals additional structure in the resonances that appear at ±0.2 V and ± 0.50 V. The probability of direct tunneling depends on the thickness of the tunnel barrier and is suppressed exponentially with increasing tunnel barrier thickness (as shown in Figure S1b), so each peak in Figure 2a (pur-ple data) could in principle represent either elastic defect-assisted tunneling, inelastic phonon-assisted tunneling, or some combination of the two.

To further characterize these resonances, two additional sets of devices are investigated. The second set of devices is constructed using hBN synthesized[38] (see the Supporting Information for detail) at Ames National Lab (blue curve; hBN thickness of 1.66±0.33 nm and area of 4.72 $\mu m^2$). While the relative amplitude of the tunneling resonances changes as compared to the commercially-supplied hBN, it is notable that the position (voltage) of many of the peaks reproduces, for example, the symmetric resonant peaks around ±0.21 V and similar peaks around -0.55 V, -0.3 V, and +0.73 V. The third set of devices is constructed with ultra-pure hBN synthesized at the International Center for Materials Nano Architectonics (MANA)[39] (green curve; hBN thickness of 2.01±0.33 nm and area of 2.09 $\mu m^2$). This material is known to yield devices with extremely high electron mobility when used to isolate, e.g., single-layer graphene from its environment, suggesting an exceptionally low level of charge-active defects[40]. This hypothesis is borne out by dI/dV measurements, which show strong suppression of resonant tunneling as compared to the first two devices (a single weak resonance is observed at -0.23 V, as shown in Figure S2).

These results confirm that the tunneling resonances observed in both commercial and Ames Lab synthesized hBN are due to intrinsic defects within the hBN material and not extrinsic impurities or contamination. As detailed in Supporting Information, this stacking process ensures a clean hBN interface without chemical exposure. Moreover, during the annealing process, thermal changes are gradual to prevent structural and functional alterations in the 2D layers caused by thermal shock. The rich variety of predicted structural defects predicted for hBN makes a definitive assignment of these resonances to specific atomic structures difficult. For example, prior work in modeling defect formation in hBN predicts the formation of native defects such as vacancies, antisites, and interstitial structural defects during hBN growth, with the most energetically favorable being carbon, oxygen, and hydrogen impurities[41–43]. Further, each defect can in principle generate multiple peaks, for example, boron vacancies (VB) have multiple different charge states: e.g., neutral, -1, and -2, with C3v symmetry for the neutral and -2 states and D3h in the -1 state[43]. Finally, the commercially supplied hBN is polycrystalline, potentially leading to varying orientations within the TJ for chemically identical defect states. However, the reproducibility across two different sets of materials synthesized in two different laboratories (commercial and Ames hBN) provides strong evidence that these resonances do in fact arise from hBN defects. In addition, the absence of the peaks in the tunneling spectra from MANA hBN further supports the absence of impurities from residues deposited during our fabrication process or other extrinsic contamination.

These measurements indicate that hBN-TJs can provide a sensitive probe of the electronic structure of atomic defects in the tunnel barrier. To further test this proposition, additional hBN crystals are prepared at Ames lab wherein carbon has been introduced during hBN growth (see Supporting Information for detail). This excess carbon can contribute to the creation of substitutional and interstitial carbon defects as well as C-nucleated and C-catalyzed defects[41,42,44]. Tunneling spectra from a device fabricated from this material are shown in Figure 2b, where an additional constellation of extremely sharp tunneling resonances is revealed. These peaks are highly reproducible, both the main figure and the inset show trace/re-trace scans of dI/dV confirming these sharp features are stable and well-defined states. Further analysis shows that individual peaks show thermal broadening consistent with tunneling spectroscopy of atomic point defects (see Figure S3).

These results demonstrate the facility with which these 2D tunnel junctions can be used to measure detailed electronic tunneling spectra for quantum point defects and impurities within a solid-state host, suggesting that this architecture may be further generalized to include encapsulated quantum systems. Specifically, we consider the molecular spin qubit VOPc[8]. Metallophthalocyanines are known molecular semiconductors, generally demonstrate environmental stability, and can be sublimed using standard evaporation techniques[8,45] resulting in well-ordered films, making them attractive targets for device integration[46,47]. To validate this potential for incorporation into the solid-state quantum devices described above, we first studied VOPc[48] (see Supporting Information for detail) thin films on a monolayer of hBN grown on highly-oriented pyrolytic graphite (HOPG) using scanning tunneling microscopy (STM). VOPc was deposited under ultra-high vacuum conditions (~1E-9 mbar)[49], with a deposition rate of ~1 monolayer/min determined by STM imaging[50]. Figure 3a and 3b show the molecular structure of VOPc and a schematic representation of the STM geometry, illustrating the heterostructure formed by VOPc deposition on hBN/HOPG. Figure 3c displays an STM topography image of a thin film of VOPc measured at 5 K, revealing the self-assembly of VOPc into a well-ordered film on the surface of the hBN/HOPG substrate. Consistent with the nominal deposition rate and prior STM studies[51], this image suggests the film comprises a uniform bilayer of VOPc[51] terminating with O-down molecules, with a low density of defects such as molecular vacancies or 3rd O-down layer molecules.

Differential conductance spectra taken at selected positions within the layer at a temperature of 5 K are shown in Figure 3d. Clear peaks in the dI/dV are resolved, representing the HOMO, LUMO and excited states of the VOPc film[52] at these distinct locations, with a weak direct tunneling background evident at higher tip-sample bias. The energies of the peaks corresponding to resonant tunneling vary with tip position, revealing spatial variation in the molecular orbitals involved in the tunneling process as shown in the spectra in Figure 3d. Furthermore, spatial variation is observed in STM images of bare hBN and dI/dV maps of the VOPc surface (refer to Figure S5). This suggests that other factors, including the moiré pattern of hBN/HOPG and local defects in both HOPG and hBN, could contribute to spatial variations in the local energy structure and density of states.

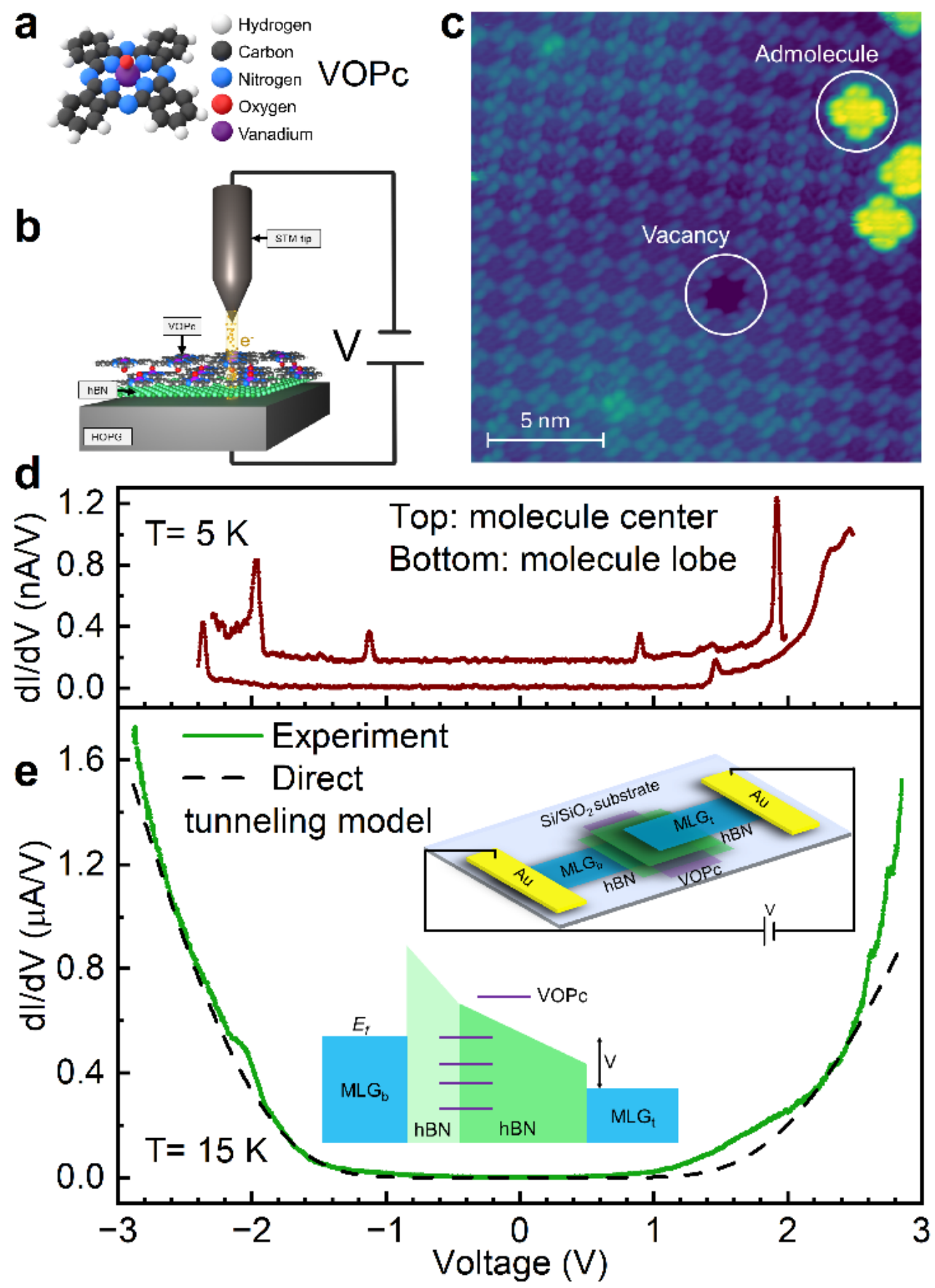

**Figure 3.** Tunneling spectroscopy of molecular qubits: VOPc on hBN. (a) Structure of VOPc molecule. (b) A schematic of the STM setup containing a 'half tunnel junction' device as studied by STM and STS (c) STM topographic image of a bilayer of VOPc deposited on a monolayer hBN on an HOPG substrate (STM imaging conditions: +2.3 V, 326 pA). (d) dI/dV of VOPc at 5 K measured by an STM setup with tip located at the molecule center (top curve) or molecule lobe (bottom curve). (e) dI/dV of VOPc encapsulated in 2D-TJ device at 15 K (green curve). The dashed black curved is direct tunneling model. Top inset, a schematic of the VOPc-TJ device with heterostructure of MLG/hBN/VOPc/hBN/MLG. Bottom inset, a schematic of the energy diagram of the device with discrete lines representing molecular states of VOPc.

The STM images indicate that VOPc thin films should form high-quality barriers in hBN TJ devices. In order to test this prediction, VOPc heterostructures were assembled in a five-layer stack (MLG/hBN/VOPc/hBN/MLG; see Supporting Information for detail). This tunneling device was then characterized by both I-V and dI/dV measurements at a temperature of 15 K (Figure 3e), as with the MLG/hBN/MLG structures in Figure 2. The dI/dV response reveals a direct tunneling background similar to the response of MLG/hBN/MLG devices fabricated from MANA hBN, but with the addition of tunneling resonances at voltages above +1 V and below -2.0 V that appear as broad peaks in the dI/dV spectrum. These voltage values are in rough agreement with the molecular states observed in STM, and also electronic transitions in the theoretical literature[17], but the peaks appear significantly broadened and a direct comparison is obscured by the relatively large direct tunneling background in the TJ device. Confirming these resonances result from VOPc molecular states, control measurements of a sample without VOPc showed only direct tunneling without any indication of resonance peaks.

To mimic the spatially averaged nature of the TJ devices, STS data were collected at different sites within the VOPc molecule and at different positions across the layer. These 21 tunneling spectra were then normalized to their maximum dI/dV signal and averaged together to generate the wine-colored curve in Figure 4. This process results in broadened peaks around -2V and + 1.5 V. To facilitate direct comparison, the direct tunneling contribution to Figure 3e was removed by fitting the curve to the Fowler-Nordheim model and subtracting it from the dI/dV response of the VOPc-TJ device (green curve, Figure 4). In this model, the direct tunneling current is given by (see Supporting Information for detail)[53]:

$$I_b = f_0 \frac{4\pi m_{eff} q}{f_1 h^3} \int_0^{\mu} E^{\frac{1}{2}} (f_1 - E)^{\frac{1}{2}} (\mu - E)\, e^{-\frac{4\kappa f_2 (f_1 - E)^{\frac{3}{2}}}{3F}} dE \qquad (1)$$

The result of this fitting process for the VOPc device is shown as a black dashed line in Figure 3e, yielding $f_1 = 9.41$ eV and $f_2 = 0.15$ for both positive and negative voltage. However, different parameters optimize the positive and negative voltage regions for $f_0$ (sample area), $f_0 = 1.6 \times 10^{-13}$ and $f_0 = 2.6 \times 10^{-13}$, respectively. The results of a similar analysis performed on the hBN devices discussed in Figure 2a are shown in Figure S6. The difference between this calculated direct-tunneling background and the experimental data is shown in Figure 4 (green curve). This residue highlights the major resonant tunneling features in the spectra, which can themselves be modeled using the following Gaussian function $\Gamma(E)$ for each resonance, with a defect current given by[29] (see Supporting Information for detail)

$$I_d = \int_{-\infty}^{\infty} \left(f(E) - f(E - F)\right)\Gamma(E)dE \quad (2)$$

This analysis yields three resonances centered at voltages of +1.6 V, -1.1 V, and -2.05 V (Figure 4, black dashed curve).

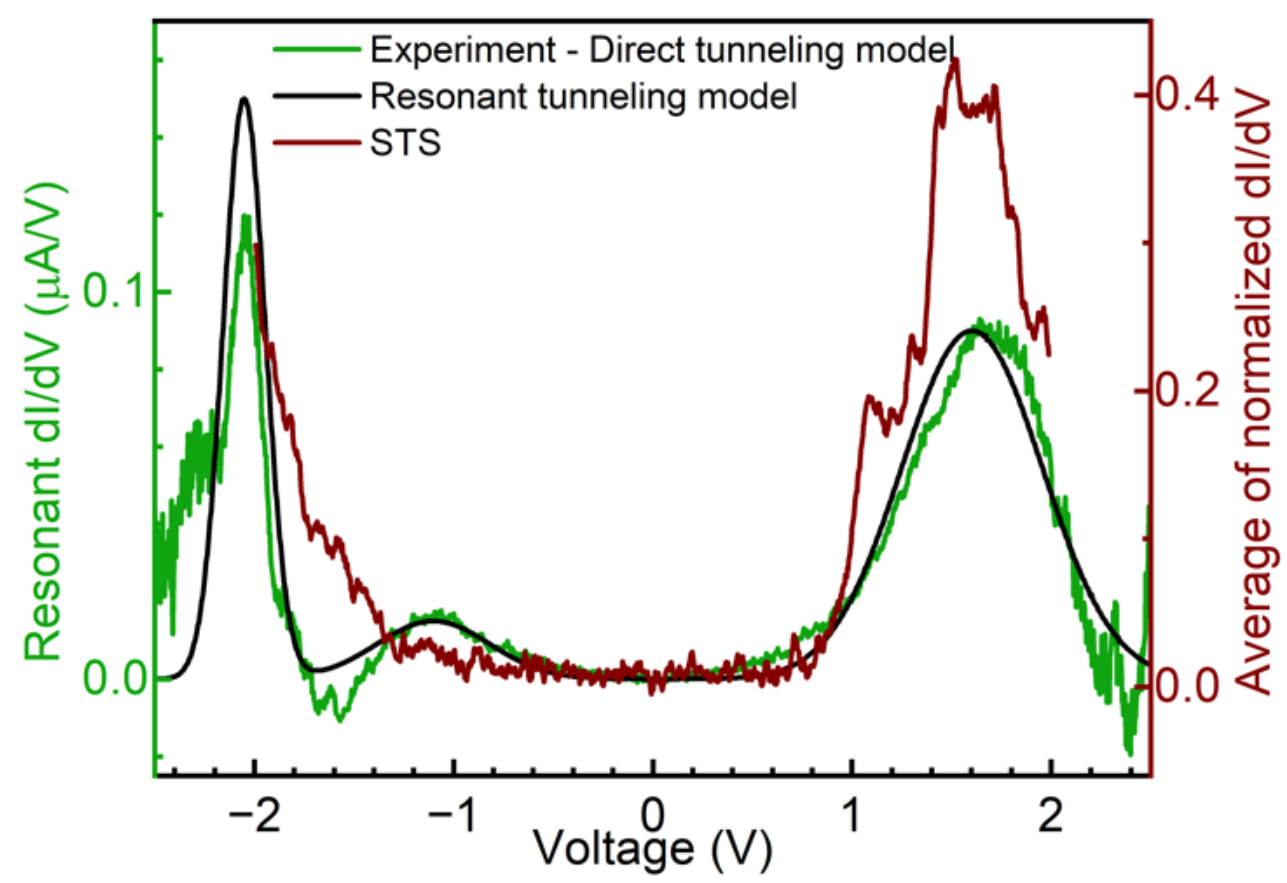

**Figure 4.** Comparison between spatially-averaged STS and VOPc devices. The green curve shows the experimental dI/dV (green curve in Figure 3e) subtracted from direct tunneling model for a VOPc-TJ device (i.e., black curve in Figure 3e). The black curve is the calculated resonant dI/dV, and the wine curve represent the average of dI/dV scans as measured by STS on a VOPc bilayer film on hBN. STS were taken at 21 different tip locations and were normalized before averaging.

Comparing the averaged STS data (wine-colored curve) with the background-subtracted tunnel junction response (green curve) reveals excellent agreement in both the gap (1.63 V) and the width of the heterogeneously broadened ensembles. This result strongly supports the hypothesis that the tunneling spectra from the tunnel junction represent a spatial average over the VOPc bilayer, with the peak width dominated by real variations in tunneling current as a function of position within the VOPc molecule and for different local configurations of the bilayer.

In summary, we have demonstrated the encapsulation of quantum point defects and molecules within a tunnel junction constructed from 2D materials, establishing a versatile and modular technique that can be applied to a wide variety of quasi-0D quantum systems. The device structure protects the QPDs and molecules during the fabrication processes, prevents direct contact between metal contacts and the 0D quantum systems (pinholes), and enables a highly sensitive readout of the electronic states of the intrin-sic defects and molecular states. We have demonstrated tunneling spectroscopy of carbon-related defects within an MLG/C-doped hBN/MLG tunnel junction device and VOPc molecular qubits encapsulated within a similar device structure (MLG/hBN/VOPc/hBN/MLG). Tunneling spectroscopy of the former shows atomic-like resonance states of the intrinsic carbon-related defects in hBN, while the latter reveals broad resonance peaks in the VOPc films that are quantitively consistent with STS of HOPG/hBN/VOPc half-devices and theory predictions of electronic structure[17]. The relative ease of fabrication and protective encapsulation demonstrates a facile approach to electronic tunneling spectroscopy for small ensembles of quasi-0D quantum systems that should be of interest for a wide variety of molecular and solid-state systems. Further, when combined with the extensive literature demonstrating high-fidelity quantum readout of electronic and nuclear spin states in related solid-state systems[31–33,54], this work provides a clear path to a dramatic expansion in the phase space of potential quantum systems to include optically inert but electrically active QPDs, molecules, and adatoms and their straightforward integration with existing solid-state quantum technologies. This device architecture enables the potential on-chip integration of microwave components for monolithic microwave-tunnel junction devices, enabling the readout and manipulation of coherent states of quantum systems[55].

ASSOCIATED CONTENT

Supporting Information

The Supporting Information is available free of charge on the ACS Publication website at DOI:

Ames hBN and VOPc syntheses, 2D stacking and device fabrication, and more details on theoretical calculations (file type: PDF)

AUTHOR INFORMATION

**Corresponding Author**

* **Ezekiel Johnston-Halperin -** Department of Physics, The Ohio State University, Columbus, Ohio, 43210, USA; Email: johnston-halperin.1@osu.edu; ORCID: 0000-0002-6240-3505

**Present Addresses**

†Daniel Pharis - Department of Physics, Cornell University, Itaca, NY 14853, USA.

†Ferdous Ara - 1800 E Paul Dirac Dr, Tallahassee, FL 32310, USA.

†Mingyu Xu - Department of Chemistry, Michigan State University, East Lansing, MI, 48824, USA.

**Author Contributions**

E. J.H. conceived the experiments and supervised the project. Z. P., M. H., M. K., E. P. H., and D. P. stacked or fabricated the 2D-TJ devices. D. E. F. and R. G. selected and produced the VOPc. Under the supervision of J. G., W. K. deposited the VOPc and with the assistance of F. A. performed the STM and STS measurements. M. K. with assistance of Z. P., M. H., and E. P. H performed the tunneling measurements of the 2D-TJ devices. M. K. analyzed the data with the assistance of W. K. in analyzing the STM data. Under the supervision of M. F., M. M. S. developed theoretical calculations for modeling direct tunneling and resonant tunneling. Under supervision of P. C., M. X. synthesized the Ames hBN. T. T. provided the MANA hBN. All authors contributed to the writing of the manuscript. All authors have given approval to the final version of the manuscript.

‡These authors contributed equally.

**Notes**

The authors declare no competing financial interest.

**ACKNOWLEDGMENT**

We acknowledge financial support from the National Science Foundation (NSF) QII-TAQS under award numbers OMA-1936219 and MPS-1936219, NSF NRT-QISE DGE-2244045, as well as the Asian Office of Aerospace Research and Development (AOARD) under award number FA238620-1-4052. Partial funding for shared facilities used in this research was provided by the Center for Emergent Materials: an NSF MRSEC under award number DMR-2011876. Work done at Ames National Laboratory was supported by the U.S. Department of Energy, Office of Basic Energy Science, Division of Materials Sciences and Engineering. Ames National Laboratory is operated for the U.S. Department of Energy by Iowa State University under Contract No. DE-AC02-07CH11358. We also thank Roland Kawakami in the Department of Physics at The Ohio State University for providing access to the transfer microscope setup in his lab, Anthony Harbo Torres in the Department of Physics at The Ohio State University for AFM analysis, Yuxin Zhang and Chun Ning Lau in the Department of Physics at The Ohio State University for providing the exfoliated MANA hBN, Zetian Mi in the Department of Electrical Engineering and Computer Science at the University of Michigan for providing hBN for STM measurements, Chhatra Tamang Lama and Yueguang Shi in the Department of Physics and Astronomy at the University of Iowa for theoretical calculations, and Joshua Kennedy at Air Force Research Laboratory (AFRL) for fruitful discussion and providing facilities for electron beam lithography.

---